\documentstyle[a4,psfig,11pt,titlepage]{article}

\newcounter{nref}
\newcommand{\bbib}{%
  \renewcommand{\refname}{\large\bf References}%
  \begin{thebibliography}{99}%
  \setcounter{enumiv}{\thenref}}
\newcommand{\ebib}{%
  \end{thebibliography}%
  \setcounter{nref}{\arabic{enumiv}}}
\newcommand{\head}[3]{%
  \setcounter{nref}{0}%
  \thispagestyle{empty}%
  \section*{\LARGE\bf #1}%
  \stepcounter{section}%
  \addcontentsline{toc}{section}{#1}%
  \large\itshape%
  #2\\\vspace{0.1pt}\\%
  #3%
  \normalsize\upshape%
  \bigskip}

\begin{document}


\head{R-Process Abundances and Cosmochronometers in Old Metal-Poor Halo 
Stars}
     {B. Pfeiffer\ $^1$, K.-L. Kratz\ $^1$, F.-K. Thielemann\ $^2$, 
      J.J. Cowan\ $^3$, C. Sneden\ $^4$, S. Burles\ $^{5,6}$, 
      D. Tytler\ $^5$, and T.C. Beers\ $^7$}
     {$^1$ Institut f\"ur Kernchemie, Universit\"at Mainz, Mainz, 
           Germany\\
      $^2$ Institut f\"ur Theoretische Physik, Universit\"at Basel, 
           Basel, Switzerland\\
      $^3$ Department of Physics and Astronomy, University of Oklahoma, 
      USA\\
      $^4$ Department of Astronomy and McDonald Observatory, University 
           of Texas, Austin, USA\\
      $^5$ Department of Physics and Center for Astrophysics and Space 
           Sciences, University of California, San Diego, USA\\
      $^6$ Department of Astronomy and Astrophysics, University of 
           Chicago, Chi\-ca\-go, USA\\
      $^7$ Department of Physics and Astronomy, Michigan State 
           University, East Lansing, USA}

\begin{figure}[ht]
   \centerline{\psfig{file=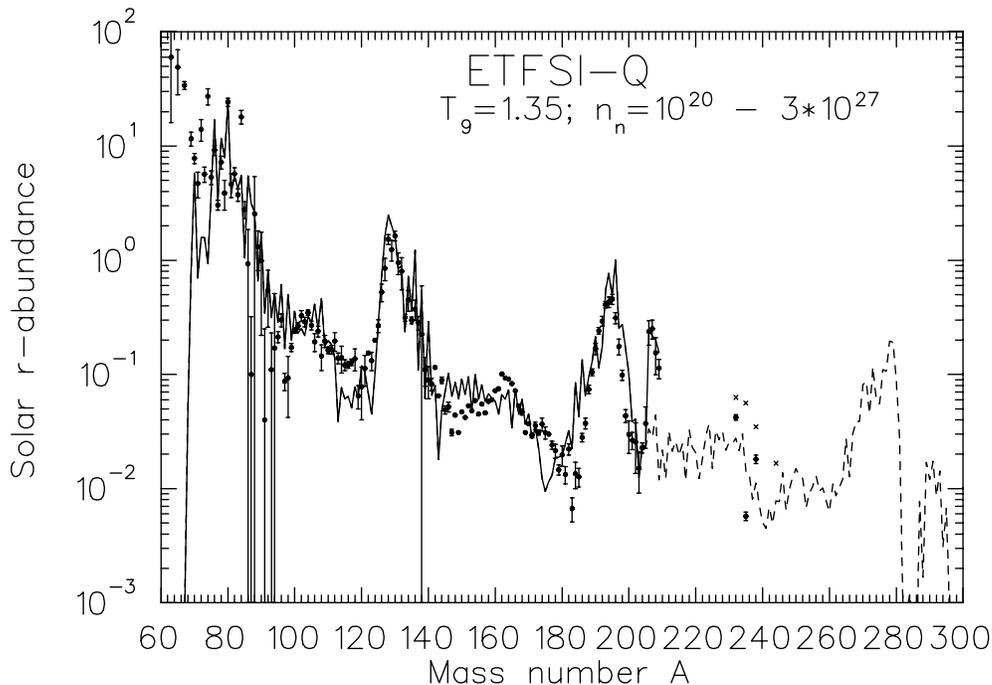,width=13.cm,angle=90}}
  \caption{Comparison of theoretical abundances with solar r-process 
abundances (small filled circles). The dashed line indicates abundances 
prior to $\beta$- and $\alpha$-decay and the solid line the final 
abundances after decay. The crosses represent calculated abundances 
after decay for the nuclei $^{232}$Th, $^{235}$U, $^{238}$U and 
$^{244}$Pu in comparison with the solar values (filled circles) for 
these nuclei.}
  \label{pfeiffer.fig1}
\end{figure}

\begin{figure}[ht]
   \centerline{\psfig{file=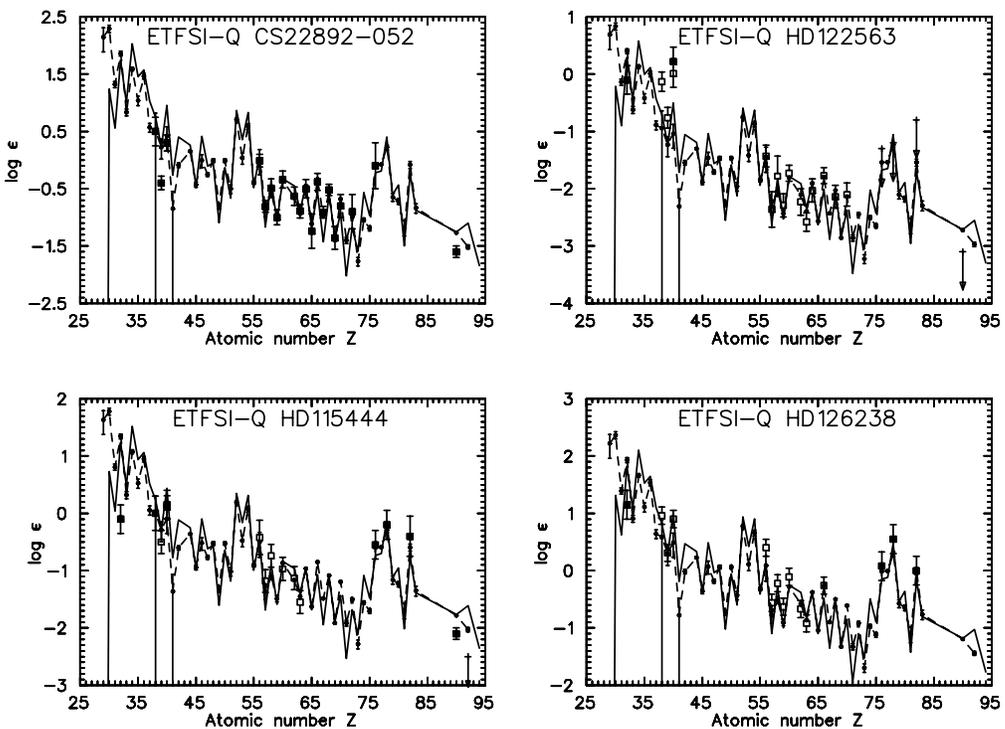,width=\textwidth,angle=90}}
  \caption{An abundance comparison between the observed neutron-capture 
elements in four metal-poor halo stars (large squares) and a theoretical 
r-process (solid line) and a solar system r-process (dashed line) 
abundance distribution.}
  \label{pfeiffer.fig2}
\end{figure}

Already 30 years ago, Seeger et al. \cite{pfeiffer.1} expressed the idea that 
the solar system r-process isotopic abundance distribution (N$_{r,\odot}$) 
is composed of several components. But only on the basis of 
new experimental and modern theoretical nuclear-physics input, Kratz et al. 
\cite{pfeiffer.2} demonstrated that within the 
so-called waiting-point approximation a minimum of three components (showing 
a steady 
flow of $\beta$-decays between magic neutron numbers) can give a reasonable 
fit to the whole N$_{r,\odot}$. A somewhat better fit can be
obtained by a more continuous superposition of exponentially declining neutron
number densities \cite{pfeiffer.3,pfeiffer.4}.
Accordingly, the s-process shows a steady flow of neutron captures in
between magic neutron numbers and a good fit is achieved when taking an 
exponentially declining superposition of exposures. Analyzing with 
present day almost perfectly known nuclear data in the s-process (as 
neutron capture and $\beta$-decay rates), Goriely \cite{pfeiffer.5} 
recently reproduced this exponential exposure with his ``multi-event'' 
model. As there is practically no experimental nuclear input in the r-process 
we prefer to apply the waiting-point approximation based on a smooth 
physical behaviour in an exponential model, rather than obtaining 
spurious results, which are just driven by obtaining a better fit with, 
however, the wrong physics. \\
Deficiencies in calculated N$_{r,\odot}$-abundances (even using the most 
recent macros\-copic-microscopic mass models FRDM and ETFSI) 
were attributed to an incorrect trend in neutron 
separation energies when approaching magic neutron numbers far from 
stability \cite{pfeiffer.2}. The weakening of shell strength 
near the neutron drip line predicted from astrophysical requirements was 
recently also obtained by 
Hartree-Fock-Bogolyubov (HFB) mass calculations with the Skyrme-P force 
\cite{pfeiffer.6}. And indeed, new spectroscopic studies of very neutron-rich 
Cd-isotopes at CERN/ISOLDE have revealed first experimental evidence for a 
quenching of the N=82 major shell below $^{132}$Sn~\cite{pfeiffer.7}.
Applying these HFB masses around the magic neutron numbers resulted in an 
eradication of the 
abundance troughs \cite{pfeiffer.4}. As large-scale HFB calculations for
deformed nuclear shapes are not yet available, Pearson et al. 
\cite{pfeiffer.8} modified their ETFSI mass model to asymptotically 
approach  the HFB masses at the drip-lines. The N$_{r,\odot}$-abundances 
calculated with these ETFS-Q masses are shown in 
Fig.~\ref{pfeiffer.fig1} to give a good fit over the whole range of 
stable r-process isotopes. This gives confidence to extrapolate the 
calculations to the unstable actinide isotopes. The abundances prior to 
$\alpha$- and $\beta$-decay are displayed in Fig.~\ref{pfeiffer.fig1} as a
dashed line and the final abundances after decay as a solid line. The 
good reproduction of the Tl-Pb-region (as endproducts of the 
$\alpha$-decay chains) let us to conclude that estimates of the initial 
abundances of the long-lived isotopes $^{232}$Th and $^{235,238}$U, which 
are applied as cosmochronometers,  can be taken from our r-process 
model. \\
Recently, stellar abundances of neutron-capture elements (beyond iron) 
have been determined over a wide Z-range in the very metal-poor Galactic 
halo star CS22892-052 \cite{pfeiffer.9}. After adjustment to solar 
metallicity, the values are consistent with the global solar-system 
r-process abundances as well as with our predictions (see 
Fig.~\ref{pfeiffer.fig2}). From this agreement, we concluded that the 
heavy elements in this star are of pure r-origin and that from the 
comparison of the observed and calculated Th/Eu abundance ratios an age 
estimate for the heavy elements of about 13~Gyr can be derived 
\cite{pfeiffer.10,pfeiffer.4}. This indicates, that r-synthesis started 
early in the Galactic evolution and that there might be a {\it {unique}} 
r-process scenario (at least beyond Z$\simeq$50). \\
As, evidently, {\bf one} single star cannot stand 
for the whole low metallicity end of the Galactic halo, further 
measurements are needed, not only to investigate other stars over a range 
of metallicities, but essentially to 
detect additional elements, especially the 3$^{rd}$-peak elements (Os, 
Pt, Pb) close to Th 
and U. These elements have absorption lines in the UV, so that they 
are best observed from space.
Therefore, spectra for three K giant stars (HD115444, HD122563, 
HD126238) were measured with the Goddard High Resolution Spectrograph on 
the Hubble Space Telescope
\cite{pfeiffer.11}. Additional high-resolution spectra were registered 
with the High Resolution Echelle Spectrometer (HIRES) at the Keck I 
telescope for the star HD115444 in particular to separate the Th absorption 
line clearly from a blending $^{13}$CH molecular line \cite{pfeiffer.11}.
The results of these observations are summarized in Fig.~\ref{pfeiffer.fig2} 
as filled squares 
together with ground-based results (open squares). After proper 
renormalization, the observed neutron-capture elements in the four stars 
displayed overlap perfectly with our theoretical r-process curve (solid line) 
and the solar system distribution (dashed line). \\
In addition to the observation of Th in CS22892-052 mentioned above, the 
new measurements yielded a firm value for HD115444 as well as an upper 
limit for HD122563. In the case of the second chronometer U, only an upper 
limit could be obtained for HD115444. The measured Th/Eu ratios combined 
with our calculated zero-age value allow to derive an 
estimate for the decay age of T=(13 $\pm$ 4)~Gyr, where the 
uncertainty takes only in account the counting 
statistics~\cite{pfeiffer.12}. This value represents a lower limit for 
the age of the Galaxy and is in line with a variety of 
recent age estimates for the Universe. \\
To summarize, the reproduction of the r-process component of solar abundances 
in the framework of the ``waiting-point approximation'' applying nuclear input 
data calculated from a macroscopic-microscopic mass model with 
Bogolyubov-enhanced shell ``quenching'' (ETFSI-Q) gives confidence in 
extrapolations beyond the stable isotopes to the actinide r-process 
cosmochronometers. The observation of ``solar'' neutron-capture element 
abundance distributions in four metal-poor halo stars indicate to a {\it 
{unique}} r-process site in the Galaxy (at least for Z$\ge$56).  This 
further strenghtens our objections to the conclusions of Goriely and 
Arnould~\cite{pfeiffer.13} that a series of several non-solar 
isotopic abundance distributions might produce a total elemental 
abundance pattern that fortuitously matches some of the neutron-capture 
elements in one low-metallicity star. Although the observation of a solar 
r-process elemental pattern is not an {\it {absolute}} proof for 
isotopic solar r-process abundances, it is nevertheless the most 
reasonable and probable conclusion.



\bbib
\bibitem{pfeiffer.1} P.A.~Seeger et~al., ApJS {\bf 11} (1965) 121.
\bibitem{pfeiffer.2} K.-L.~Kratz et~al., ApJ {\bf 403} (1993) 216. 
\bibitem{pfeiffer.3} B.~Chen et~al., Phys. Lett. {\bf B355} (1995) 37.
\bibitem{pfeiffer.4} B.~Pfeiffer et~al., Z. Phys. {\bf A357} (1997) 235.
\bibitem{pfeiffer.5} S.~Goriely, A\&A {\bf 327} (1997) 845.
\bibitem{pfeiffer.6} J.~Dobaczewski et~al., Phys. Scr. {\bf T56} (1995) 
15.
\bibitem{pfeiffer.7} K.-L.~Kratz et~al., Proc. Int. Conf. on ``Fission and 
Properties of Neutron-Rich Nuclei'', Sanibel Island, 1997, World 
Scientific Press, in print
\bibitem{pfeiffer.8} J.M.~Pearson et~al., Phys. Lett. {\bf B387} (1996) 
455.
\bibitem{pfeiffer.9} C.~Sneden et~al., ApJ {\bf 467} (1996) 819.
\bibitem{pfeiffer.10} J.J.~Cowan et~al., ApJ {\bf 480} (1997) 246.
\bibitem{pfeiffer.11} C.~Sneden et~al., ApJ {\bf 496} (1998) 235.
\bibitem{pfeiffer.12} J.J.~Cowan et~al., submitted to ApJ.
\bibitem{pfeiffer.13} S.~Goriely and M.~Arnould, A\&A {\bf 322} (1997) L29.
\ebib
\end{document}